\documentclass[12pt]{article}
\begin{document}
\begin{center}
{\Large \bf Final state interaction  in $K\rightarrow 2\pi$ decay }\\
\vspace{5mm} {{\bf E.P. Shabalin}\footnote{{\it E-mail address:}
shabalin@itep.ru (E.Shabalin)}} \\{\it Institute for Theoretical
and Experimental Physics, B.Cheremushkinskaya, 25, 117218, Moscow,
Russia }\\ \end{center} \vspace{1cm}
\noindent{\bf Abstract}\\
{\small Contrary to wide-spread opinion that the
final state interaction (FSI) enhances the amplitude $<2\pi;I=0|K^0>$, we
argue that FSI does not increase the absolute value of this
amplitude.}\\

\vspace{3mm}

\vspace{1cm}

The essential progress in understanding the nature of the $\Delta I=1/2$
rule in $K\to 2\pi$ decays was achieved in the paper \cite{1}, where
the authors had found a considerable increase of contribution of the
operators containing a product of the left-handed and right-handed quark
currents generated by the diagrams called later the penguin ones.  But for
a quantitative agreement with the experimental data, a search for some
additional enhancement of the $<2\pi;I=0|K^0>$ amplitude produced by
long-distance effects was utterly desirable. A necessity of additional
enhancement of this amplitude due to long-distance strong interactions was
also noted later in \cite{2}.

The attempts to take into account the long-distance effects were undertaken
in \cite{3} - \cite{15}.

In \cite{3}, the necessary increase of the amplitude $<2\pi;I=0|K^0>$ was
associated with 1/N corrections calculated within the large-N approach
(N being the number of colours).

In \cite{4}, \cite{5}, the strengthening of the $<2\pi;I=0|K>$ amplitude
arised due to a small mass of the intermediate scalar $\sigma$ meson.

One more mechanism of enhancement of the $<2\pi;I=0|K^0>$ amplitude was
ascribed to the final state interaction of the pions \cite{6} - \cite{14}.
But as it will be shown in present paper,
unitarization of the $ K\to
2\pi$ amplitude in presence of FSI leads to the opposite effect: a
decrease of the $<2\pi;I=0|K^0>$ amplitude. The similar conclusion had
been obtained formerly in \cite{15} where a different from our approach
to investigation of FSI effect was used.
We exploit the technique based on the effective $\Delta S =1$ non-leptonic
Lagrangian \cite{1} \begin{equation} L^{\rm{weak}} =\sqrt{2}G_F \sin
\theta_C \cos \theta_C \sum_i c_i O_i.
\end{equation}
Here $O_i$ are the four-quark operators and $c_i$ are the Wilson
coefficients calculated taking into account renormalization effect
produced by strong quark-gluon interaction at short distances.
Using also the recipe for bosonization of the diquark compositions
proposed in \cite{2}, one obtains the following result:
\begin{equation}
<\pi^+(p_+),\pi^-(p_-);I=0|K^0(q)>=\kappa^{(0)}(q^2-p^2_-),
\end{equation}
where
$\kappa^{(0)}$ is a function of $G_F, F_{\pi}, \theta_C$ and some
combination of $c_i$. The numerical values of $\kappa^{(0)}$ obtained in
\cite{1} and \cite{2} turned out to be insufficient for a reproduction of
the observed magnitude of the $<2\pi;I=0|K^0>$ amplitude.

Could a rescattering the final pions occuring at long distances change
the situation? To answer this question, we consider at first the elastic
$\pi\pi$ scattering itself.\\
\noindent{\bf The elastic $\pi\pi$ scattering.} \\
The general form of the amplitude of elastic $\pi\pi $ scattering is
\begin{equation}
T=<\pi_k(p'_1) \pi_l(p'_2)|\pi_i(p_1)\pi_j(p_2)> =A\delta_{ij} \delta_{kl}
+B\delta_{ik}\delta_{jl} + C\delta_{il}\delta_{jk},
\end{equation}
where $k,l,i,j$ are the isotopical indices and $A,B,C$ are the functions
of $s=(p_1+p_2)^2$, $t=(p_1-p'_1)^2$ ,$u=(p_1-p'_2)^2$.

The amplitudes with the fixed isospin $I=0,1,2$ are
\begin{equation}
T^{(0)}=3A+B+C, \qquad T^{(1)}=B-C, \qquad T^{(2)}=B+C .
\end{equation}
To understand the problems arising in description of $\pi\pi$ scattering in
the framework of field theory, let's consider the simplest chiral $\sigma$
model, where
\begin{equation}
A^{\rm{tree}}=
\frac{g^2_{\sigma\pi\pi}}{m^2_{\sigma}-s}-\frac{g^2_{\sigma\pi\pi}}
{m^2_{\sigma}-m^2_{\pi}}
=\frac{g^2_{\sigma\pi\pi}}{m^2_{\sigma}-m^2_{\pi}}\cdot
\frac{s-m^2_{\pi}}{m^2_{\sigma}-s}
\end{equation}
and $B$ and $C$ are obtained from $A$ by replacement $s\to t$ and $s\to
u$, respectively.

It follows from Eqs.(4) and (5), that the isosinglet amplitude
$T^{(0)}_{\rm{tree}}$ is a sum of the resonance part \begin{equation}
A_{\rm{Res}}^{\rm{tree}}=3A^{\rm{tree}} \end{equation} and the
potential part \begin{equation} A^{\rm{tree}}_{\rm{Pot}}=
B^{\rm{tree}}+C^{\rm{tree}}. \end{equation} The resonance part must be
unitarized summing up the chains of pion loops, that is , taking into account the repeated
rescattering of the final pions.

At the one loop order
\begin{equation}
A^{\rm{one-loop}}_{\rm{Res}}=A^{\rm{tree}}_{\rm{Res}}(1+\Re\Pi_R+i
\Im\Pi)=A^{\rm{tree}}_{\rm{Res}}(1+ \Re\Pi_R+
i\frac{A^{\rm{tree}}_{\rm{Res}}\sqrt{1-4m^2_{\pi}/s}}{16\pi}),
\end{equation}
where $\Re \Pi_R$ is the renormalized real part of the closed pion loop
\cite{16}
\begin{equation}
\Re\Pi_R(s)=\Re\Pi(s)-\Re\Pi(m^2_{\sigma}) -\frac{\partial
\Re\Pi(s)}{\partial s}|_{s=m^2_{\sigma}}(s-m^2_{\sigma}).
\end{equation}
The last two terms in r.h.s. of this equation are absorbed in
renormalization of the resonance mass and coupling constant
$g_{\sigma\pi\pi}$. In leading order of perturbation theory in
$g_{\sigma\pi\pi}$ one has (see also \cite{17})
\begin{equation}
\Re \Pi_R(s)=\frac{A^{tree}_{res}(s)}{16
\pi^2}\sqrt{1-4m^2_{\pi}/s} \ln\frac{1-\sqrt{1-4m^2{\pi}/s}}{1+\sqrt{1-
4m^2_{\pi}/s}}.
\end{equation}
But in view of very big value
of this constant such a calculation does not give a proper estimate of
$\Re \Pi _R(s)$. It will be explained below, how to get a reliable
magnitude of $\Re\Pi_R(s)$.

The unitarized expression for $A_{\rm{Res}}$ is \footnote{Strictly
speaking, the $4\pi$ intermediate state brings a correction in Eq.(11).
But its contribution to $\Im \Pi(s)$ is equal to zero because
$4m_{\pi}>m_K$. As for $\Re \Pi_R(s)$, in our approach, all separate
contributions to it will be taken into account phenomenologically
introducing a form factor, see below.}

\begin{equation}
A^{\rm{unitar}}_{\rm{Res}}=\frac{A^{\rm{tree}}_{\rm{Res}}(s)}{1-\Re\Pi_R(s)-i\Im\Pi_{\rm{Res}}}=
\frac{A^{\rm{tree}}_{\rm{Res}}(s)}{1-\Re\Pi_R(s)}\cdot \frac{1}{1-i\tan
\delta_{\rm{Res}}},
\end{equation}
where
\begin{equation}
\tan \delta_{\rm{Res}}=\frac{A^{\rm{tree}}_{\rm{Res}}(s)
\sqrt{1-4m^2_{\pi}/s}}{16\pi(1-\Re\Pi_R(s))}.
\end{equation}
The Eq.(11) may be rewritten in the form
\begin{equation}
A^{\rm{unitar}}_{\rm{Res}}=\frac{16\pi \sin \delta_{\rm{Res}}
e^{i\delta_{\rm{Res}}}}{\sqrt{1- 4m^2_{\pi}/s}}, \end{equation} leading to
the cross section \begin{equation} \sigma_{\rm{Res}}=\frac{4\pi
\sin^2\delta_{\rm{Res}}}{k^2}, \qquad k=\frac{\sqrt{s}}{2} \cdot
\sqrt{1-4m^2_{\pi}/s}.  \end{equation}

Of course, the amplitude $T^{(0)}$ must be unitarized including the
potential part $B+C$ too. But if this potential part is considerably
smaller than the resonance one, the effect of FSI can be estimated
roughly from $A^{\rm{unitar}}_{\rm{Res}}$. To
understand what gives the unitarization of $A^{\rm{tree}}_{\rm{Pot}}$
, we use the form of the $S$ matrix of elastic
scattering with the total phase shift as a
sum of the phase shifts produced by separate mechanisms of scattering
\cite{18}. In other words, if there is a number of resonances and if, in
addition, there is potential scattering, the matrix $S$ looks as
\begin{equation} S=e^{2i\delta_{\rm{Res1}}}
e^{2i\delta_{\rm{Res2}}}...e^{2i\delta_{\rm{Pot}}}. \end{equation} Then,
in terms of \begin{equation} \delta_{\rm{Res}}=\sum_j \delta_{\rm{Resj}}
\quad\mbox{and} \quad
\delta_{\rm{tot}}=\delta_{\rm{Res}}+\delta_{\rm{Pot}}, \end{equation}
\begin{equation} A^{\rm{unitar}}=\frac{16\pi}{\sqrt{1-4m^2_{\pi}/s}}\sin
\delta_{\rm{tot}}e^{i\delta_{\rm{tot}}}
\end{equation}
or
\begin{equation}
A^{\rm{unitar}}=\frac{16\pi}{\sqrt{1-4m^2_{\pi}/s}}(\sin \delta_{\rm{Res}}
\cos \delta_{\rm{Pot}}+\sin \delta_{\rm{Pot}} \cos
\delta_{\rm{Res}})e^{i\delta_{\rm{tot}}}. \end{equation}

The phase shifts $\delta_{\rm{Res}}$ and $\delta_{\rm{Pot}}$ can be taken
from \cite{19}, where the Resonance Chiral Theory of $\pi \pi$ Scattering
(RChT) was elaborated. This model incorporates two $\sigma$ mesons,
$f_0(980)$, $\rho(750)$ and $f_2(1270)$. In addition, some
phenomenological form factors were introduced in the vertices
$\sigma\pi\pi, \rho\pi\pi, f_2\pi\pi$.  Their appearence follows in the
field theory from the result (11), according to which the effect of $\Re
\Pi_R(s)$ may be incorporated in $g^2_{\sigma \pi\pi}(s)$, where
\begin{equation} g^2_{\sigma \pi\pi}(s)=\frac{g^2_{\sigma \pi\pi}}{1-\Re
\Pi_R(s)}= g^2_{\sigma \pi\pi}F(s).  \end{equation} The RChT gives a
quite satisfactory description of the observed behavior of the phase
shifts $\delta^0_0(s)$, $\delta^2_0(s)$, $\delta^1_1(s)$ in the range
$4m^2_{\pi}\le s \le 1\; \mbox{(GeV)}^2$. For $\delta^0_0$ and
$\delta^2_0$ at $s=m^2_K$ we have \begin{equation}
\delta^0_0=36.74^{\circ}, \qquad \delta^2_0=-8.1^{\circ} \end{equation}
in agreement with the experimental values $\delta^0_0=37^{\circ}\pm
3^{\circ}, \quad \delta^2_0=-7^{\circ} \pm 1^{\circ}$ (see \cite{14}).
The phase shifts $\delta^0_2(s)$ and $\delta^2_2(s)$ turn out to be
consistent with the results obtained using the Roy's dispersion relations.

A good fit to experimental data on $\delta^0_0$ and $\delta^2_0$ was
obtained in \cite{19} with
\begin{equation}
F(q^2)=exp\left(-0.5(q^2-m^2_{\pi})/(\mbox{GeV})^2 \right).
\end{equation}
Then
\begin{equation}
\Re \Pi_R(s=m^2_K)=-0.12
\end{equation}
and has the same sign as in Eq.(10), but turns out to be considerably
smaller in absolute value. For $\sqrt{s}=m_K$, the phase shifts obtained
in \cite{19} are\footnote{A difference $1.3^{\circ}$ between
$\delta_{Pot}$ and $\delta^2_0$ is caused by intermediate $\rho$ meson
giving the contributions to these phase shifts which are different in value
and sign \cite{19}.}

\begin{equation}
\delta_{\rm{Res}}=46.71^{\circ}, \qquad \delta_{\rm{Pot}}=-9.40^{\circ}.
\end{equation} Then \begin{equation}
\frac{A^{\rm{unitar}}_{\rm{Pot}}}{A^{\rm{unitar}}_{\rm{Res}}}=\frac{\sin
\delta_{\rm{Pot}} \cos \delta_{\rm{Res}}}{\sin \delta_{\rm{Res}} \cos
\delta_{\rm{Pot}}}=-0.156.  \end{equation} Therefore, the amplitude
$A^{\rm{unitar}}_{\rm{Pot}}$ is small and may be neglected in a rough
estimate of FSI effect.\\
\noindent {\bf FSI in $K^0 \to 2\pi$ decay.}

Basing on the result (24), we shall estimate effects of FSI
in the $K \to 2\pi$ amplitude, taking into account only
the resonance rescattering effect.  Then, in one loop approximation,
the amplitude (2) is \begin{equation}
\begin{array}{lll} <\pi^+ (p_+)
\pi^-(p_-);I=0|K^0(q)>^{\rm{one-loop}}_{\rm{Res}}=    \\ \kappa^{(0)}
\left[(q^2-p^2_-)+ \frac{A^{\rm{tree}}_{\rm{Res}}(q^2)}{(2\pi)^4
i}\int\frac{(q^2-p^2)d^n
p}{[(p-q)^2-m^2_{\pi}][p^2-m^2_{\pi}]}+i\frac{A^{\rm{tree}}_{\rm{Res}}(q^2)}{16\pi}
(q^2-p^2_-)\sqrt{1-
4m^2_{\pi}/q^2}\right].
\end{array}
\end{equation}

In the t'Hooft-Veltman scheme of dimensional regularization \cite{20}
\begin{equation}
\begin{array}{lll}
\frac{1}{(2\pi)^4i}\int\frac{d^n p}{[(p-q)^2-m^2][p^2-m^2]}=
\frac{1}{16\pi^2}\left(\ln\frac{M^2_0}{m^2}+2+  \sqrt{1-4m^2/q^2} \ln
\frac{1- \sqrt{1-4m^2/q^2}}{1+\sqrt{1-4m^2/q^2}} \right);\\
\frac{1}{(2\pi)^4}\int \frac{p^2 d^np}{[(p-q)^2-m^2][p^2-m^2]}=
\frac{m^2}{16\pi^2}\left(2\ln\frac{M^2_0}{m^2}+3+\sqrt{1-4m^2/q^2}\ln
\frac{1-\sqrt{1-4m^2/q^2}}{1+\sqrt{1-4m^2/q^2}} \right)\\
M_0\to \infty .
\end{array}
\end{equation}
After renormalization excluding the parts of these integrals independent
of the external momentum, we come to
\begin{equation}
\begin{array}{lll}
<\pi^+ \pi^-|K^0(q)>^{\rm{one-loop}}_{\rm{on-mass-shell}}= \\
=\kappa
(m^2_K-m^2_{\pi})\left[1+\frac{A^{\rm{tree}}_{\rm{Res}}(s)}{16\pi^2}\sqrt{1-
4m^2_{\pi}/s} \ln \frac{1-\sqrt{1-4m^2_{\pi}/s}}{1+\sqrt{1-4m^2_{\pi}/s}} +
i\frac{A^{\rm{tree}}_{\rm{Res}}(s)}{16\pi}\sqrt{1-4m^2_{\pi}/s} \right].
\end{array}
\end{equation}
This result agrees with the Cabibbo--Gell-Mann theorem \cite{21}, according
to which the $K\to 2\pi$ amplitude vanishes in the limit of exact
$SU(3)$ symmetry.

The unitarization of the amplitude (27) done
in accordance with the prescription (11) leads to the result
\begin{equation}
|<\pi\pi;I=0|K^0(q^2=m^2_K)>|_{\rm{Res}}=\kappa^{(0)}
(m^2_K-m^2_{\pi})\frac{\cos\delta_{\rm{Res}}}{1-\Re
\Pi_R(m^2_K) }, \end{equation} where $\Re \Pi_R(m^2_K)$ is determined by
Eq.(22). In accordance with Eq.(12), the factor
characterising FSI effect can be expressed through $\delta_{\rm{Res}}$ and
$A^{\rm{tree}}_{\rm{Res}}$: \begin{equation}
\chi^{(0)}=\frac{\cos\delta_{\rm{Res}}(s=m^2_K)}{1-\Re\Pi_R(m^2_K)}
=\frac{16\pi \sin
\delta_{\rm{Res}}(s=m^2_K)}{A^{\rm{tree}}_{\rm{Res}}(s=m^2_K)\sqrt{1-
4m^2_{\pi}/m^2_K}}.
\end{equation}
The phase shift $\delta_{\rm{Res}}$ obtained in the framework
of the theory \cite{19} and given by Eq.(23) is compatible with the
data giving  $\delta^0_{\rm{Res}}\approx(\delta^0_0
-\delta^2_0)^{\rm{exp}}=44.0^{\circ} \pm 3.2^{\circ}$.

A value of $A^{\rm{tree}}_{\rm{Res}}$ evidently depends on the used
$\sigma$ model which, besides a good description of data on the phase
shifts, must reproduce the other experimental observations. In particular,
the result for a mass of the lightest $\sigma$ meson: $m_{\sigma} \le 700
$ MeV \cite{22}. An existance of $\sigma$ particle with mass $m_{\sigma}=
600\div 700$ MeV explains also why the observed value of the form factor
$F_1(Q^2= 4m^2_{\pi}) $ of $K_{e4}$ decay is by 50\% bigger than a value
predicted in the non-physical point $Q^2=m^2_{\pi}$ by the theory based on
algebra of currents and soft-pion technics \cite{23} \footnote{In the
region $ m^2_{\pi}\ge Q^2 \le 4m^2_{\pi}$ the amplitude is purely real}.

Using the formulae represented in \cite{23} and in Appendix in \cite{24}
for the coupling constants and masses of $\sigma$ mesons in
$U(3)_L\otimes U(3)_R$ symmetric meson theory, one can obtain that at
$m_{\sigma} \le 700$ MeV the normalized amplitude \begin{equation}
A^{\rm{tree}}_{\rm{Res}}= \frac{3}{2}
(s-m^2_{\pi})\left[\frac{g^2_{\sigma_1\pi\pi}}{ (m^2_{\sigma_1
}-m^2_{\pi})(m^2_{\sigma_1 }-s)}
+\frac{g^2_{\sigma_2\pi\pi}}{(m^2_{\sigma_2}-m^2_{\pi})
(m^2_{\sigma_2}-s)}\right] \ge 72.
\end{equation}
In the theory \cite{19} $m_{\sigma_1}=697.6$ MeV and
$A^{\rm{tree}}_{\rm{Res}}=72.4$.  Therefore, using the data on the phase
shifts, one obtains \begin{equation} \chi^{(0)}=0.585^{+0.033}_{-0.024}.
\end{equation} This value is compatible with $\chi^{(0)}=0.61$ given by
Eq.(28) where $\Re \Pi_R(s=m^2_K)=-0.12$. The part connected with the
potential rescattering, being small, can not change the conclusion that FSI
diminishes the $<2\pi;I=O|K^0>$ amplitude.

In the case of the $<2\pi;I=2|K^0>$ amplitude, FSI adds to the initial
expression $\kappa^{(2)}(m^2_K-m^2_{\pi})$ the factor
\begin{equation}
\chi^{(2)}=16\pi \sin \delta^2_0 e^{i\delta^2_0}/(A^{\rm{tree}}_2
\sqrt{1-4m^2_{\pi}/m^2_K})
\end{equation}
At the used in \cite{19} parameters, the normalized
$A^{\rm{tree}}_2 =-8.293, \qquad
\delta^2_0=-8.1^{\circ} \quad \mbox{and as a result}\quad
\chi^{(2)}=1.035$.

The influence of FSI on the $K^0\to 2\pi$ decay was studied in the
framework of $\sigma$ model in \cite{25}. In this paper, the
authors, however, put $\Re \Pi=0$. Then
$$
A^{\rm{unitar}}=A^{\rm{tree}}/(1-i\Im \Pi)
$$
and this formula was used by them to estimate the FSI effects in the $K^0
\to 2\pi$ decay. But earlier the same authors had found that $\Re \Pi \ne
0$ \cite{17}. In this case, the unitarization leads to
$A^{\rm{unitar}}$ for the elastic $\pi\pi$ scattering given in
Eq.(11)  and to $A^{\rm{unitar}}_{\rm{Res}}(K\to 2\pi; I=0)$
in Eq.(28). As it is seen from Eq.(28), FSI could increase
or diminish the $K\to 2\pi$ amplitude depending on relative magnitudes of
$\cos \delta$ and $(1-\Re \Pi_R)$ . We have shown that
$\cos\delta/(1-\Re\Pi_R)<1$, that allows us to affirm that FSI
diminishes the isosinglet part of the $K \to 2\pi$ amplitude.

\noindent{\bf Conclusion.}\\
Carring out our investigation in the framework of field theory with the
phenomenological mesonic Lagrangian we
have not found an enhancement of the amplitude $<\pi\pi;I=0|K^0>$ due to
final state interaction of pions.  On the contrary, our analysis has shown
that FSI diminishes this amplitude.  Hence, FSI is not at all the
mechanism bringing us nearer to explanation of the $\Delta I=1/2$ rule in
the $K\to 2\pi$ decay.\\

\vspace{1cm}

\noindent{\bf Acknowledgements}\\

\vspace{5mm}

I am very grateful to Yu.A.Simonov for discussions and comments concerning
this letter.


\vspace{5mm}

\begin{thebibliography}{99}{\small \bibitem{1}  A.I.
Vainshtein, V.I.  Zakharov, M.A. Shifman, Zh. Eksp. Teor. Fiz. (JETP) 72
(1977) 1275; [ Sov.  Phys. JETP. 45 (1977) 670 ];\\  M.A. Shifman, A.I.
Vainshtein, V.I.  Zakharov, Nucl. Phys. B 210 (1977) 316.  \bibitem{2}
W.A. Bardeen, A.J.  Buras, J.-M. Gerard, Nucl. Phys. B 293 (1987) 787.
\bibitem{3} W.A.  Bardeen, A.J. Buras, J.-M. Gerard, Phys. Lett. B 192
(1987) 138.  \bibitem{4}
E.Shabalin, Yad. Fiz. 48 (1988) 272; [ Sov. J. Nucl. Phys.
 48 (1988)].
\bibitem{5} T. Morozumi, C.S.  Lim, A.I.Sanda,
Phys. Rev. Lett. 65 (1990) 404; \\  Y.- Y. Keum, U.  Nierste, A.I.
Sanda, Phys. Lett. B 457 (1999) 157.  \bibitem{6} M.P. Losher,
V.E.  Markushin, H.O. Zheng, Phys. Rev. D 55 (1997) 2894.
\bibitem{7} E.A.  Paschos, hep-ph/9912230. \bibitem{8} E.  Pallante, A.
Pich, Phys. Rev. Lett. 84 (2000) 2568. \bibitem{9} E. Pallante,
A. Pich, Nucl. Phys. B 592 (2001) 294. \bibitem{10} A.A. Bel'kov,
et al., Phys. Lett. B 220 (1989) 459.  \bibitem{11} N.
Isgur, et al., Phys. Rev. Lett. 64 (1990) 161. \bibitem{12}
G.E.  Brown, et al., Phys. Lett. B 238 (1990) 20.
\bibitem{13} M.  Neubert, B. Stech, Phys. Rev. D 44 (1991) 775.
\bibitem{14} S.  Bertolini, J.O.  Eeg, M. Fabbrichesi, Rev. Mod. Phys.
72 (2000) 65.
\bibitem{15} T.N. Truong, Phys.Lett. B 207 (1988) 495.
\bibitem{16} S.S.
Schweber, H.A. Bethe, F.de Hoffman, Mesons and Fields, Eds.: Row, Peterson
and Company, Evanston,Illinois, 1955.  \bibitem{17} N.N. Achasov, S.A.
Devyanin, G.N. Shestakov, Yad. Fiz. 32 (1980) 1098 ;
[Sov.J.Nucl.Phys.  {\bf 32} (1980)]; Preprint TPh 109, Novosibirsk 1980;
Phys. Lett. B 96 (1980) 168; Uspekhi Fiz. Nauk, 142 (1984) 361.
\bibitem{18} L.D. Landau, E.M. Lifshitz, Quantum Mechanics
(Non-relativistic Theory),  Chapter XVIII, FizMatGiz, 1963.\\ S.Ishida
et al., Prog. Theor. Phys. 95 (1996) 745.  \bibitem{19} E.P.
Shabalin, Phys. At. Nucl. 63 (2000) 594.  \bibitem{20} G.
t'Hooft, Nucl. Phys. B 62 (1973) 444; \\  G. t'Hooft, M.
Veltman, Nucl. Phys. B 44 (1972) 189. \bibitem{21}  N.  Cabibbo,
Phys. Rev. Lett., 12 (1964) 62; \\ M.  Gell-Mann, Phys. Rev. Lett.,
12 (1964) 155.
\bibitem{22} Proc.of Workshop "Possible Existance of the $\sigma$-meson
and Its Implication to Hadron Physics", Kyoto, Japan, June 12-14, 2000.
KEK Proceedigs 2000-4. (Eds. S. Isida et al.).
\bibitem{23}  E.P. Shabalin, Yad.Fiz. 49 (1989) 588 [Sov.J.Nucl.Phys. 49
(1989) 365].
\bibitem{24}  E.P. Shabalin, Nucl.Phys. B 409 (1993) 87.
\bibitem{25} N.N. Achasov, G.N. Shestakov, Phys. Rev. D 49
(1994) 5779.}




\end{thebibliography}
\end{document}